\documentclass[prl,aps,twocolumn,groupedaddress,floats,showpacs,final,superscriptaddress]{revtex4}
\usepackage{graphicx}
\usepackage{dcolumn}
\usepackage{bm}
\usepackage{color}
\definecolor{blue}{rgb}{0.3,0.3,0.9}

\usepackage[pdftex]{hyperref} 

\begin{document}

\author{Zhiyuan Yao}
\affiliation{Department of Physics, University of Massachusetts, Amherst, MA 01003, USA}
\author{Karine P. C. da Costa}
\affiliation{Department of Physics, University of Massachusetts, Amherst, MA 01003, USA}
\affiliation{Instituto de F\'{\i}sica, Universidade de S\~{a}o Paulo, 05508-090, S\~{a}o Paulo, Brazil}
\author{Mikhail Kiselev}
\affiliation{The Abdus Salam International Centre for Theoretical Physics, Strada Costiera 11, I-34151 Trieste, Italy}
\author{Nikolay Prokof'ev}
\affiliation{Department of Physics, University of Massachusetts, Amherst, MA 01003, USA}
\affiliation{Russian Research Center ``Kurchatov Institute'', 123182 Moscow, Russia}


\title{Critical Exponents of the Superfluid-Bose Glass Transition in Three-Dimensions}
\date{\today}


\begin{abstract}
Recent experimental and numerical studies of the critical-temperature exponent $\phi$ for
the superfluid-Bose glass universality in three-dimensional systems report strong violations
of the key quantum critical relation, $\phi=\nu z$, where $z$ and $\nu$ are the dynamic
and correlation length exponents, respectively, and question the conventional
scaling laws for this quantum critical point.
Using Monte Carlo simulations of the disordered Bose-Hubbard
model, we demonstrate that previous work on the superfluid-to-normal fluid
transition-temperature dependence on chemical potential (or magnetic field, in spin systems),
$T_c \propto (\mu-\mu_c)^{\phi}$, was misinterpreting {\it transient} behavior on approach to
the fluctuation region with the genuine critical law. When the model parameters are modified
to have a broad quantum critical region, simulations of both quantum and classical models
reveal that the $\phi=\nu z$ law [with $\phi=2.7(2)$, $z=3$, and $\nu = 0.88(5)$]
holds true, resolving the $\phi$-exponent ``crisis".
\end{abstract}


\maketitle


Disordered Bose-Hubbard (DBH) model is frequently employed as a key prototype system
to discuss and understand a number of important experimental cases, such as $^{4}$He in porous media and on various substrates, thin superconducting films, cold atoms in disordered optical lattice potentials, and disordered magnets (see \cite{Yu1,Zhelud} and references therein), \emph{etc}.

The pioneering work \cite{Giamarchi,Fisher} on the DBH model has established that at $T=0$
an insulating Bose glass (BG) phase will emerge as a result of localization effects
in disordered potentials.
On a lattice, this phase will intervene between the Mott-insulator (MI) and
superfluid (SF) phases at arbitrary weak disorder strength \cite{Fisher,theorem} and
completely destroy the MI phase at strong disorder. In contrast with the gapped incompressible
MI phase, the BG phase has finite compressibility, $\kappa$, due to finite density
of localized gapless quasiparticle and quasihole excitations.
Using scaling arguments, and the fact that
$\kappa =const$ at the critical point of the quantum SF-BG transition, it was predicted that the dynamic critical exponent, $z$, always equals the dimension of space; i.e., $z=d$ \cite{Fisher}. The decrease of the normal-to-superfluid transition temperature,
$T_c$, on approach to the quantum critical point (QCP) is characterized by the $\phi$ exponent:
$T_c \propto (g_c-g)^\phi$,  where $g$ is the control parameter used to reach the QCP.
Standard scaling analysis of the quantum-critical free-energy density predicts
that $\phi$ has to satisfy the relation $\phi=\nu z$.  Therefore, taking into account Harris criterion $\nu \ge 2/d$ \cite{Harris} for the correlation length exponent in disordered systems,
it is expected that $\phi \ge 2$, within the standard picture of quantum critical phenomena.

Despite substantial research efforts in the last two decades, some aspects of the universal
critical behavior described above remain controversial (see, e.g., Ref.~\cite{Zheludev}).
For instance, Ref.~\cite{Weichman} argues that finite $\kappa$ at the SF-BG critical point might come
from the regular analytic (rather than singular critical) part of the free energy, and, thus,
$z<d$ should be considered as an undetermined critical exponent. Moreover, recent
experiments on magnetic systems \cite{Yu1}, as well as quantum Monte Carlo simulations
of related disordered $S=1$ antiferromagnets with single-ion anisotropy \cite{Yu2}, which
use magnetic field (equivalent to the chemical potential in the bosonic system) as a control parameter to drive the system to quantum criticality, report compelling evidence that
the values $\phi \approx 1.1(1)$ and $\nu \approx 0.75(10)$ are in strong violation
of the key relation $\phi=z\nu $ and the bound $\phi \ge 2$.
As a result, finite-temperature scaling relations used to describe SF-BG
criticality for decades, are challenged.

In this Letter, we address the $\phi$-exponent ``crisis'' in the three-dimensional SF-BG universality class by performing accurate studies of quantum and classical models using Monte Carlo simulations based on Worm Algorithm \cite{Worm,easyworm} and established protocols of measuring critical points
using finite-size scaling (FSS) plots of mean-square winding number fluctuations (see, e.g., Ref.~\cite{soyler}) averaged over disorder realizations (typically 5000-20000 realizations).
With regard to previous studies, we find that they were performed
away from the quantum critical region, and the genuine critical behavior was simply out of reach---the transition temperature drops below the detection limit before the data become suitable for extraction of $\phi$. However, the low-$T_c$ problem is avoided when the SF-BG transition is approached by increasing disorder strength at constant particle density. In this regime,
simulations of the $(d+1)$-dimensional classical \emph{J}-current model
(in the same universality class) reveal that $z=d=3$, $\phi = 2.7(2)$, $\nu =0.88(5)$
are fully consistent with the $\phi=\nu z$ relation. This conclusion is further confirmed by
quantum Monte Carlo simulations of the hard-core DBH, putting an end to the controversy.

\begin{figure}[htbp]
\includegraphics[scale=0.4,angle=0,width=0.7\columnwidth]{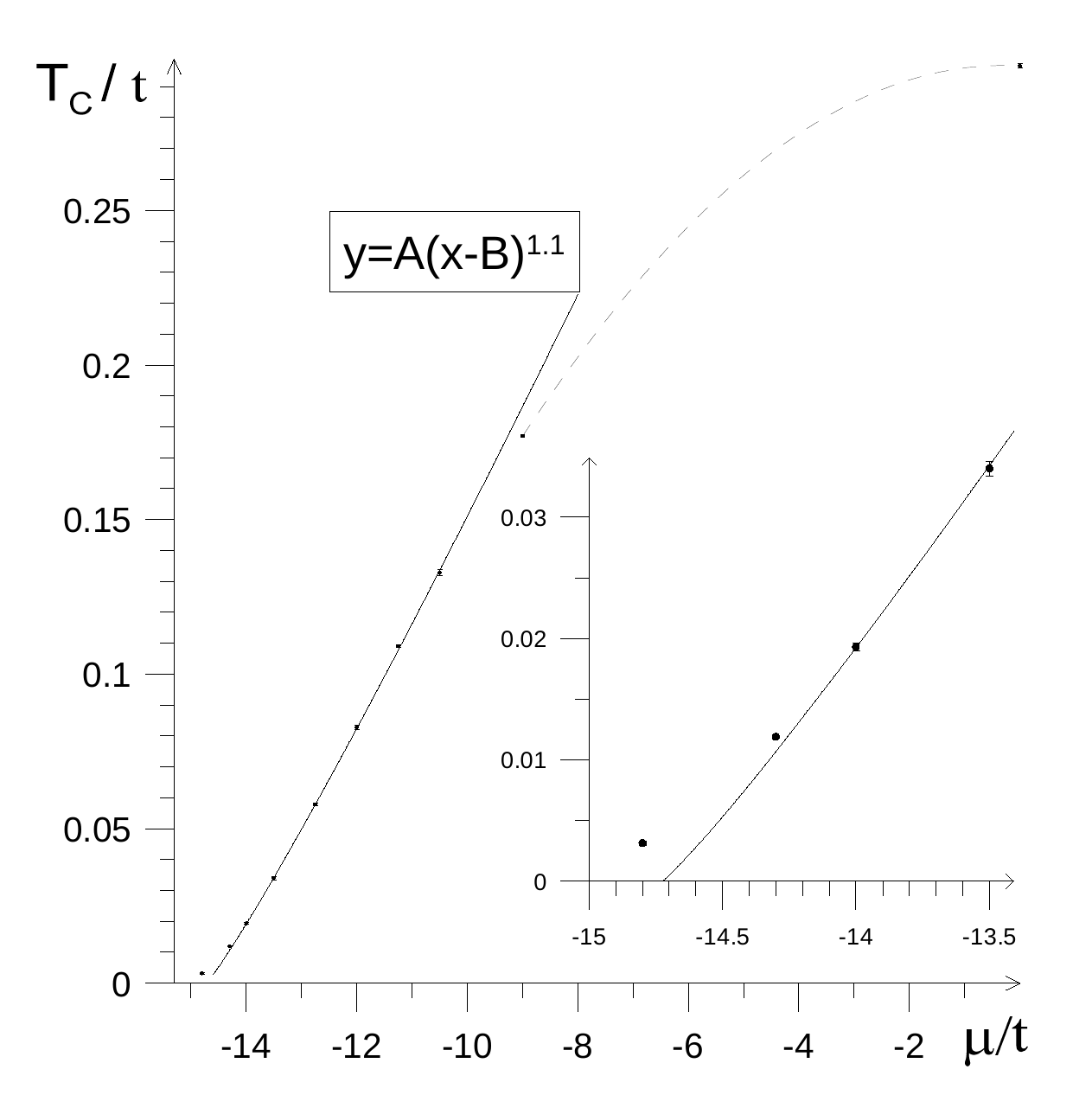}
\caption{\label{fig:1}  Critical temperature of the hard-core Bose-Hubbard model as a function of chemical potential for disorder strength $\Delta/t=16$ fitted to the $T_c=A(\mu-\mu_c)^{1.1}$
power law. The dashed line is to guide an eye. }
\end{figure}

\begin{figure}[htbp]
\includegraphics[scale=0.4,angle=0,width=0.7\columnwidth]{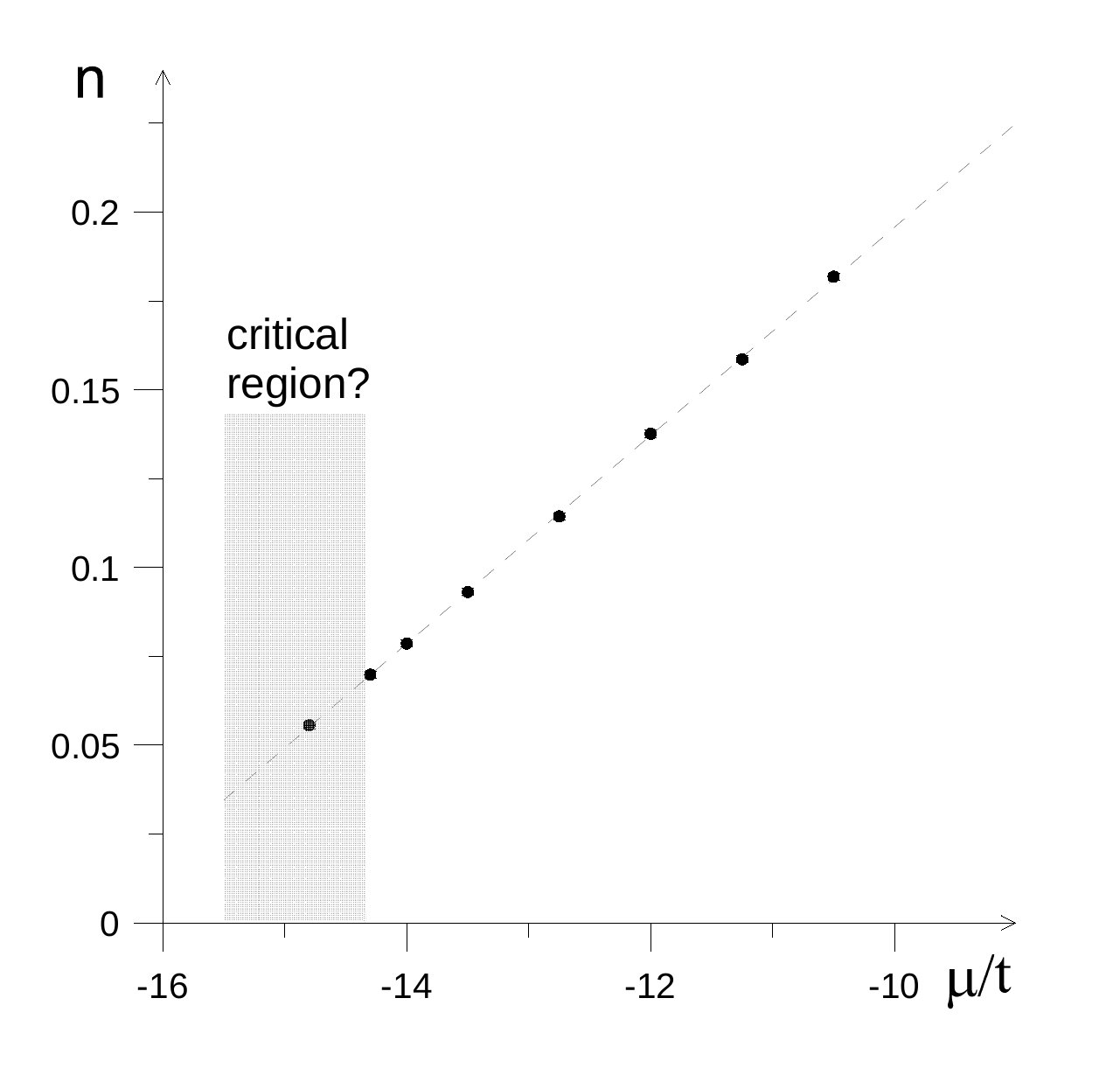}
\caption{\label{fig:2} Density at the thermal critical point of model (\ref{eq:disBH})
as a function of chemical potential for $\Delta/t=16$. The dashed line is a linear fit.}
\end{figure}

Consider the hard-core DBH on the simple cubic lattice
(equivalent to the spin-1/2 $XY$-ferromagnet in magnetic field) with the Hamiltonian
\begin{equation} \label{eq:disBH}
H=-t \sum_{\langle ij \rangle} \left( a_{i}^{\dagger}a_{j}^{\,}+ h.c. \right) -\sum_i \mu_i n_i \;,
\end{equation}
where $a_{j}$ is the bosonic annihilation operator, $t$ is the hopping amplitude,
$n_i=a_{i}^{\dagger} a_{i}^{\,}$ is the particle number operator with the hard-core constraint
$n_i \le 1$, ${\langle \cdots \rangle}$ stands for summation over the nearest-neighbor sites,
and $\mu_i=\mu+\delta \mu_i$. Here $\mu$ is the chemical potential and $\delta \mu_i$ is a bounded random potential with uniform distribution on the $[-\Delta,\Delta]$ interval and un-correlated in space. The SF-BG transition is induced by fixing disorder strength at $\Delta/t=16$ and
decreasing the chemical potential, similarly to the protocol employed
in Refs.~\cite{Yu1,Yu2,Zheludev}. Our data for $T_c(\mu )$ are shown in Fig.~\ref{fig:1}.
They feature an extended region in the parameter space where $T_c(\mu)$ is decreasing
by closely following the reported $(\mu-\mu_c)^{1.1}$ law.
However, with highly accurate data for $T_c$  (our system sizes are at least
an order of magnitude larger than in previous work) we observe that the last point
is deviating from this power-law well outside of its error bar, see inset in Fig.~\ref{fig:1},
indicating that most of the points in Fig.~\ref{fig:1} might not be in the critical regime yet.
This observation is confirmed by revealing the $n(\mu )$ dependence in Fig.~\ref{fig:2}.
Since density remains finite at the QCP, one requirement of being in the quantum critical
region is to have $n(\mu )-n(\mu_c)\ll n(\mu_c)$. This condition is clearly violated for most of the points used to establish the $T_c\propto (\mu-\mu_c)^{1.1}$ law in previous studies at low fields.


Since current problems with scaling relations are likely originating from strong $n(\mu)$ dependence when $\mu$ is used as a control parameter (leading to the critical region with extremely small $T_c$ values), we radically change the strategy and study the SF-BG criticality as a function of disorder strength $\Delta$ at constant density. Universal properties of QCPs in $d$-dimensions can
be equally well studied using $(d+1)$-dimensional classical mappings which are algorithmically superior from the numerical point of view. The simplest classical counterpart of the hard-core DBH in $d=3$ is the $(3+1)$-dimensional \emph{J}-current model \cite{Jcurrent}
\begin{equation}
\beta H = K\sum_{n,\alpha}  J_{n,\alpha}^2 -\sum_{n}\mu_{\vec{r}} J_{n,\tau} \;,
\label{jcurrent}
\end{equation}
with the $J_{n,\alpha = \tau} = 0,1$ and $J_{n,\alpha \ne \tau } = -1,0,1$ constraints.
Here index $\alpha$ enumerates space-time directions  $\hat{x}, \hat{y}, \hat{z}, \hat{\tau}$, $n=(\vec{r},\tau)$ is the site index in the hyper-cubic
space-time lattice, $\mu_{\vec{r}}=\mu+\delta{\mu_{\vec{r}}}$ is the chemical potential plus bounded random potential energy that depends on space coordinate only.
The random potential $\delta{\mu_{\vec{r}}}$ is uncorrelated in space and is uniformly
distributed on the $[-\Delta,\Delta]$ interval. An integer valued current $J_{n,\alpha}$ is
defined on lattice bonds $\langle n,n+\alpha \rangle$ and satisfies the divergence-free condition;
i.e., $\sum_{\alpha} \left( J_{n,\alpha} + J_{n,-\alpha} \right) = 0$, where it is understood that $J_{n,-\alpha}=-J_{n-\alpha,\alpha}$. Graphically, the configuration space
is composed of \emph{J}-current loops mimicking path-integral trajectories of bosonic particles.
In terms of the underlying bosonic system, $\{ J_{n,\alpha = \tau} \}$ and $\{ J_{n, \alpha \ne \tau}$ 
represent the on-site occupation
numbers and hopping transitions, respectively, while $K \propto 1/t$.

\begin{figure}[htbp]
\includegraphics[scale=0.4,angle=0,width=0.8\columnwidth]{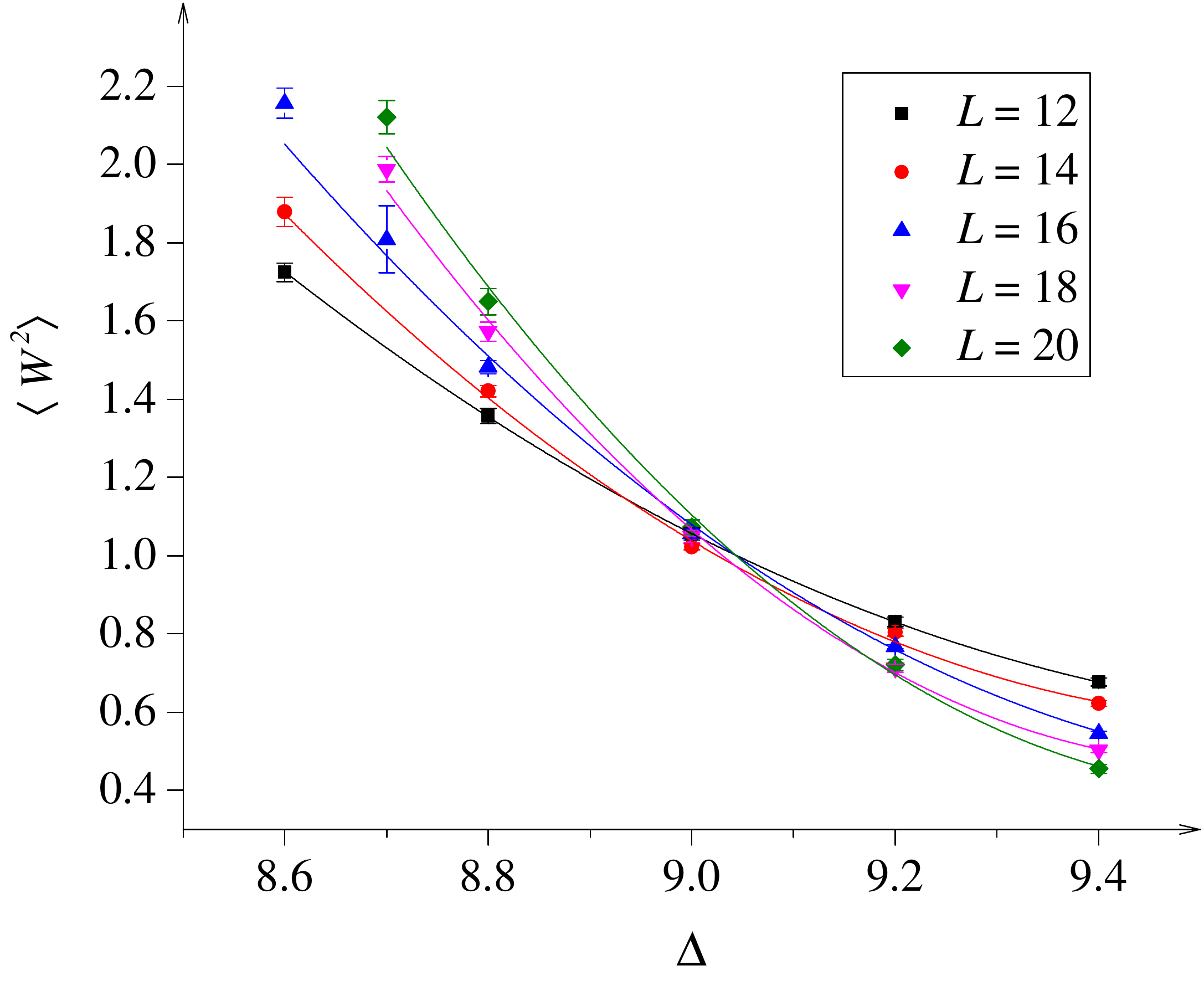}
\caption{\label{fig:3} (Color online.)
Finite-size scaling plots for $\langle W^2 \rangle = \tilde{f}(L^{1/\nu} \delta )$ for system sizes $L=12$ (black), $L=14$
(red), $L=16$ (blue), $L=18$ (magenta), and $L=20$ (green) with fixed ratio $L_{\tau}
 =2L^3$. Data points are fitted with second-order polynomials. We do not observe corrections
to scaling within our error bars.}
\end{figure}

Accurate determination of the critical exponent $\phi$ ultimately rests on precise location
of the QCP, or critical disorder strength $\Delta_c$, where the power law originates.
[Otherwise, one can be easily
mislead by the transient behavior (similarly to one shown in  Fig.~\ref{fig:1}). Likewise,
all data points for the \emph{J}-current model can be fit nearly perfectly with the power law
based on $\phi \approx 3.3$ if $\Delta_c$ is kept as a free parameter.]
To determine $\Delta_c$ along with the correlation length exponent $\nu$, we employ
FSS of scale-invariant mean-square winding number fluctuations,
\begin{equation}
\left\langle W^2\right\rangle =(1/d)\sum_{\alpha=x,y,z}
\left \langle W^2_{\alpha} \right\rangle \;,
\end{equation}
where $W_{\alpha}=(1/L_{\alpha}) \sum_n J_{n,\alpha}$ is the winding number in $\alpha$ direction.
If small detuning from the QCP is characterized by $\delta=(\Delta_c-\Delta)/\Delta_c$, then
the correlation lengths in space and time directions, $\xi$ and $\xi_\tau$, diverge as $\xi_{\tau} \propto \xi ^z \propto |\delta|^{-\nu z}$, and $\langle W^2 \rangle$ is a universal function
of length scale ratios
\begin{equation} \label{eq:w2}
\langle W^2 \rangle = f ( L/\xi,L_{\tau}/\xi_{\tau} )=\tilde{f} ( L^{1/\nu} \delta )\;.
\end{equation}
In the last equality we assume that the ratio $L_{\tau}/L^z$ is fixed.
By plotting $\langle W^2 \rangle$ for different system sizes, one determines the critical parameter
from the crossing point of $\tilde{f}$ curves (if $z$ was guessed correctly).
We argue that $z=d$ is an exact relation. Indeed, in the vicinity of QCP the compressibility can be formally decomposed into critical and regular (non-singular) parts $\kappa(\Delta) = \kappa_{s}(\delta) + \kappa_{reg}(\delta )$ with  $\kappa_{s} \propto |\delta|^{\nu (d-z)}$ \cite{Fisher}.
One may speculate that finite $\kappa(\delta=0)$ is due to regular part, while the critical part vanishes at $\delta=0$.  However, this possibility is immediately ruled out by observation
that finite $\kappa$ in the BG phase is due to localized single-particle modes,
while such modes do not exist in the superfluid phase. Thus, finite $\kappa(0)$
is entirely due to critical modes and $z=d$ (our FSS data are in perfect agreement
with this conclusion, see Fig. \ref{fig:3}).

Our simulations of model (\ref{jcurrent}) were done with $K=2$ at half-integer filling
factor, when $\mu=K$. For FSS at the QCP we fix $L_{\tau}/L^3=2$ and consider only
large system sizes from $N=2\times 12^6$ to $N=2\times 20^6$ sites (we hit the limit
of what modern computer cluster can handle in reasonable time, given that every parameter point
has to be averaged over $5000-20000$ disorder realizations). The crossing of $\tilde{f}$-curves
shown in Fig. \ref{fig:3} pinpoints the critical disorder strength to be at
$\Delta_c=9.02(5)$.

\begin{figure}[htbp]
\includegraphics[scale=0.4,angle=0,width=0.8\columnwidth]{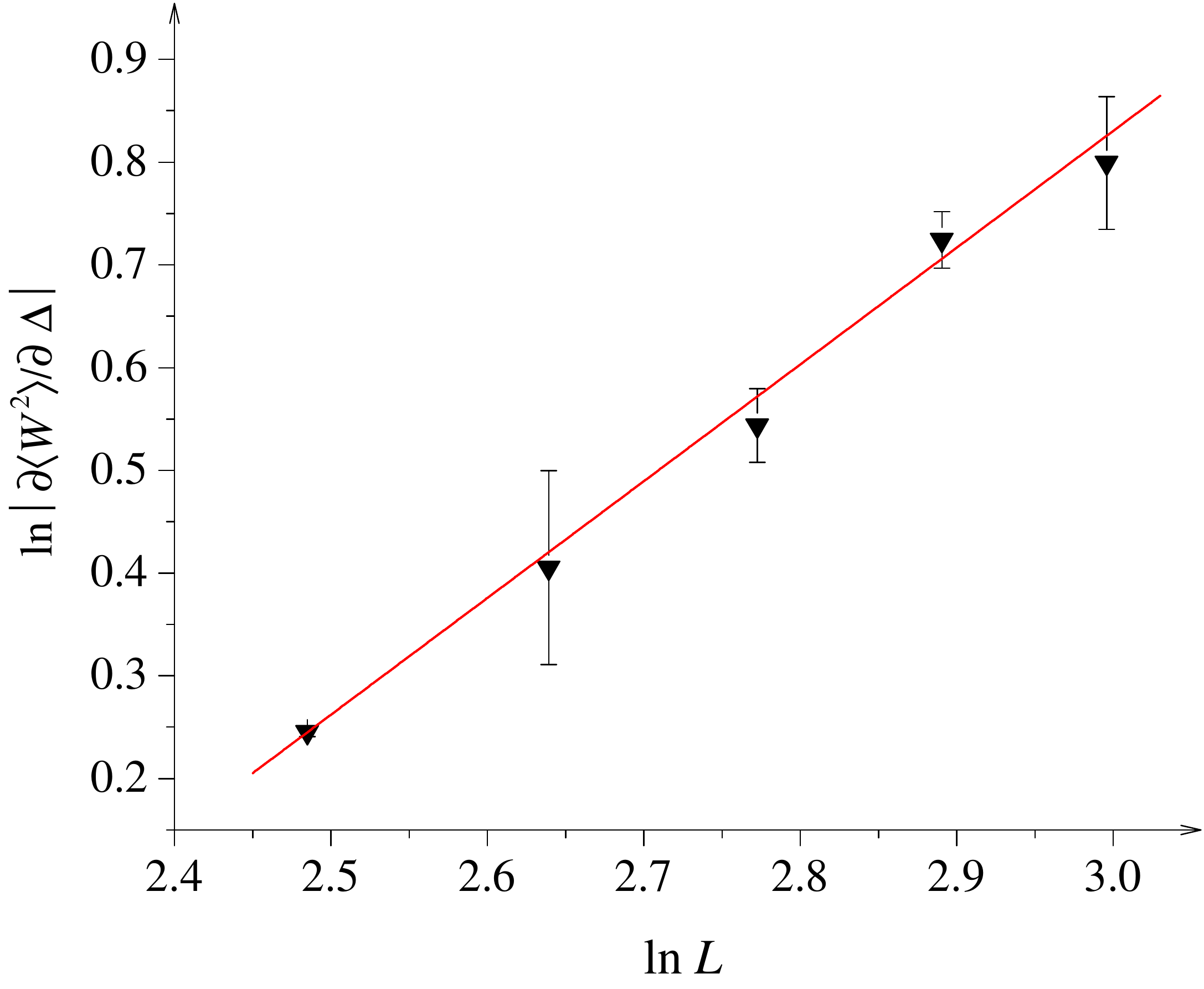}
\caption{\label{fig:4}
Deducing $1/\nu$ from the linear fit of
$\ln | \partial \langle W^2 \rangle / \partial \Delta |$ as a function of
$\ln L$ using 4 points near the critical point, $\Delta=8.8, 9.0, 9.2, 9.4.$
Error bars are based on the uncertainty of the fitting procedure, given the data points
and their statistical error bars in Fig.~\ref{fig:3}.}
\end{figure}

From Eq.~(\ref{eq:w2}), it follows that at the critical point
\begin{equation}
\partial \langle W^2 \rangle / \partial \Delta = const \times L^{1/\nu} \;,
\end{equation}
enabling one to determine the correlation length exponent $\nu$ from the slopes of
universal curves at the crossing point. The corresponding analysis is shown
in Fig. \ref{fig:4} where $\nu=0.88(5)$ is deduced from the log-log plot
of $\tilde{f}$ derivatives.
This result is in full agreement with previous findings \cite{Yu2,sorensen}.

\begin{figure}[htbp]
\includegraphics[scale=0.4,angle=0,width=0.8\columnwidth]{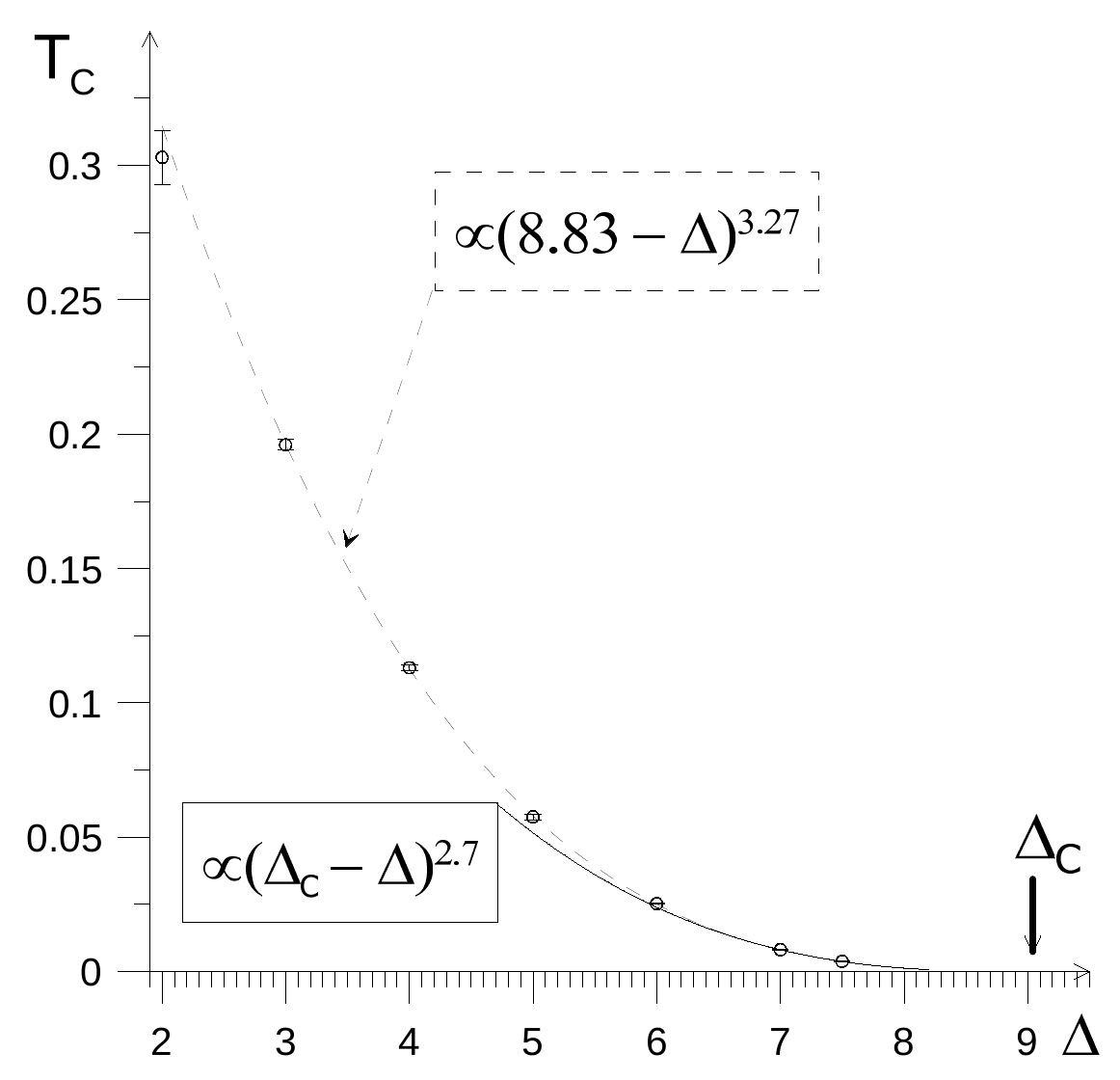}
\caption{\label{fig:5} Critical temperature of the \emph{J}-current model as a function of disorder strength. Solid line is the power-law fit to the lowest transition temperatures
assuming known location of the quantum critical point. Dashed line is a power-law originating
from $\Delta = 8.83$.}
\end{figure}

We now proceed to the evaluation of the critical-temperature exponent $\phi$ from
accurate measurements of $T_c(\Delta)$ (using similar FSS analysis)
and the power-law $ T_c = A\delta^{\phi}$ fit to the lowest transition temperatures,
see Fig.~\ref{fig:5}.
In striking contrast to Fig.~\ref{fig:1} and previously reported results \cite{Yu1,Yu2},
all data points nicely follow the power-law curve $T_c \propto (8.83-\Delta)^{3.27}$
as $T_c$ decreases nearly two orders in magnitude! If $\Delta_c$ were left undetermined we
would have to conclude that $\phi \approx 3.3$. However, if the power-law fit is performed
with the known value of QCP (i.e., with $\Delta_c=9.02$), the prediction is different:
The $\phi$ exponent decreases from $2.9$ to $2.7$ as we reduce the number of the
lowest-temperature points to be included in the fit from $T_c < 0.1$ to $T_c < 0.01$.
We thus claim our final result as $\phi =2.7(2)$, which is in good agreement
with the prediction based on the quantum critical relation $\phi=z\nu$ with $z=3$ and
$\nu=0.88(5)$. [The order parameter exponent deduced from the constant-density
approach, $\beta = 1.5(2)$, also differs significantly
from the value $\beta \approx 0.6(1)$ characteristic of
the transient $\mu/t \ge -14$ interval.]

\begin{figure}[htbp]
\includegraphics[scale=0.4,angle=0,width=0.8\columnwidth]{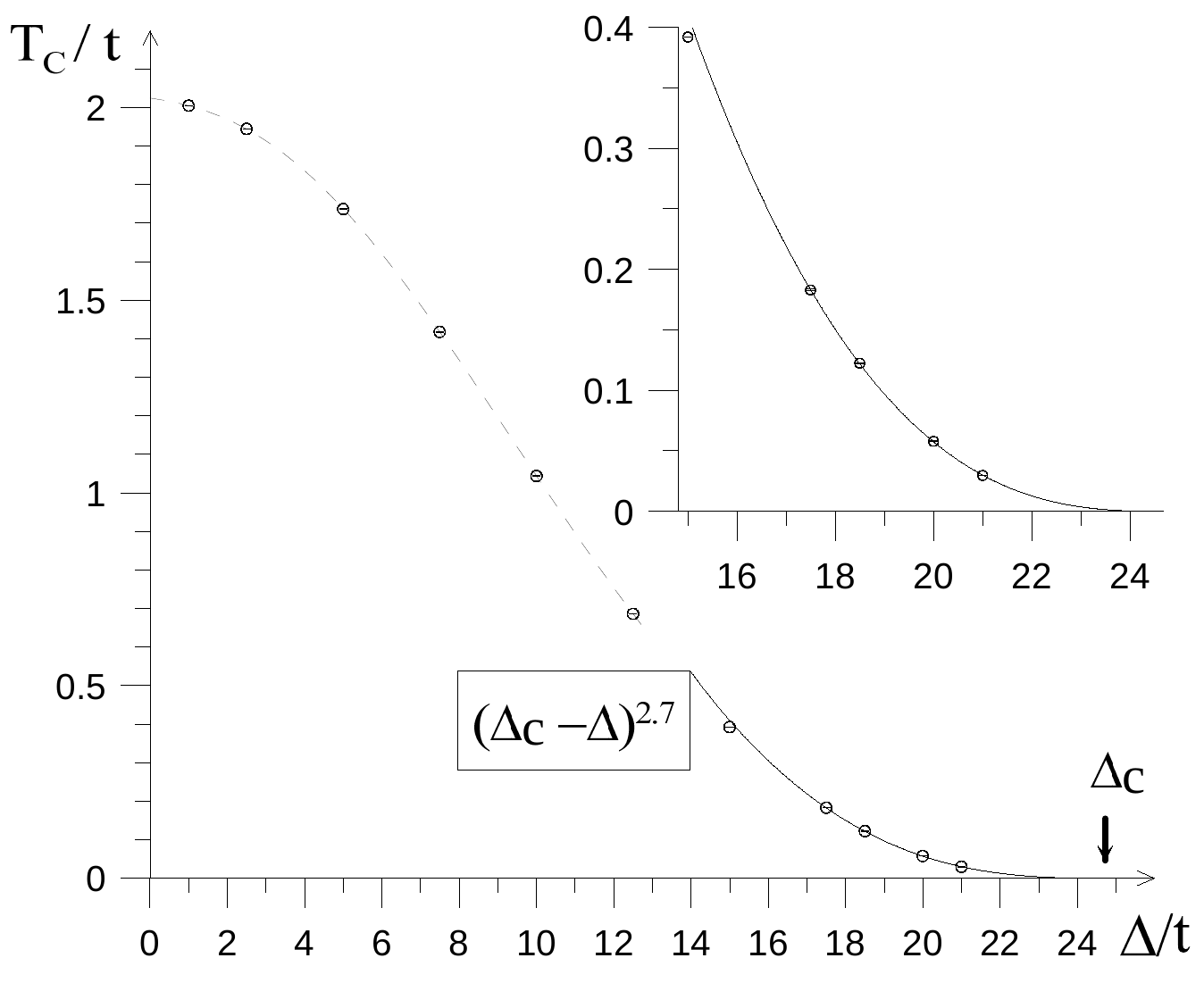}
\caption{\label{fig:6} Critical temperature dependence on disorder strength
in the hard-core DBH at half-integer filling factor. The solid line is a fit of the last
five points to the $A(\Delta_c-\Delta)^\phi$ law with exponent $\phi=2.7$ fixed
at the value determined from simulations of the \emph{J}-current model.
From this fit we predict that the quantum critical point is located at
$\Delta_c \approx 24.67$. Error bars are shown but are smaller than the symbol size.
Inset: Zoom in to the tail of the main plot.}
\end{figure}

To verify the universality of our findings and to shed light on what to expect if a similar
study is attempted experimentally using magnetic or cold-atom systems, we performed
quantum Monte Carlo
simulation of model (\ref{eq:disBH}) at half-integer filling factor (i.e., at $\mu=0$, or zero external magnetic field in the case of spin-1/2 $XY$-ferromagnet). Our data for normal-to-superfluid
transition temperature as a function of disorder strength are shown in Fig.~\ref{fig:6}
($T_c (\Delta)$ was determined from FSS analysis of $\langle W^2 \rangle$ plots with $8\le L \le 64$).
Given that simulations of quantum models are more challenging numerically, we did not attempt
to determine $\Delta_c$ and averaged results over smaller number of disorder realizations, from
$5000$ at high temperature to $500$ at low temperature. The lowest transition temperatures
can be perfectly fitted to the $T_c \propto (\Delta_c-\Delta)^{2.7}$ law with
$\Delta_c/t=24.67$. This critical behavior starts at temperatures as high as $T_c/t <0.5$ and
we were able to verify it down to $T_c/t \approx 0.03$, see Fig.~\ref{fig:6} inset.
There is no doubt that the $\phi >2$ condition is satisfied at the SF-BG transition.

In summary, we addressed the current $\phi$-exponent ``crisis" for the superfluid-to-Bose Glass
universality class in three dimensions. Previous work questioned conventional scaling relations
$z=d$ and $\phi=z\nu$ with $\nu>d/2$ for the SF-BG quantum critical point.
Using extensive Monte Carlo simulations of the hard-core DBH and its
classical \emph{J}-current counterpart we were able to identify problems with previous analysis
(strong dependence of density/magnetization on chemical potential/external magnetic field on approach to quantum criticality). We argued that $z=d$ is an exact relation, and used
it to determine the critical-temperature exponent $\phi$ from simulations of the
\emph{J}-current model. Our final result $\phi=2.7(2)$ is in good agreement with the
quantum critical prediction $\phi = z \nu = d\nu $ based on $\nu=0.88(5)$, putting the
controversy to an end. We verified universality of our findings and determined under
what conditions the $\phi$ exponent can be studied experimentally.

We thank Y. Deng for help with simulations. M. K. appreciates fruitful discussions with A. Zheludev. This work was supported by the National Science Foundation under the grant PHY-1314735, the MURI Program ``New Quantum Phases of Matter" from AFOSR; the work of K. P. C. da C. was  supported by FAPESP. We also thank ICTP (Trieste), the Aspen Center for Physics and the NSF Grant \# 1066293 for hospitality during the crucial stages of this work.


\end{document}